\documentclass[lineno]{jfm}

\usepackage{graphicx}
\usepackage{newtxtext}
\usepackage{newtxmath}
\usepackage{natbib}
\usepackage{hyperref}
\hypersetup{
    colorlinks = true,
    urlcolor   = blue,
    citecolor  = black,
}

\newcommand{\RomanNumeralCaps}[1]

\title{Unification of The Navier-Stokes Equation for Fluid Mechanics and Aeroacoustics}

\author{Tapan K. Sengupta\aff{1}
  \corresp{\email{tksen@iitk.ac.in}}
\affiliation{\aff{1} Professor (Retired), High Performance Computing Laboratory, IIT Kanpur, UP-208 016, India}
}

\begin{document}
\maketitle

\begin{abstract}
The Navier-Stokes equation for a compressible Newtonian fluid is reassessed, in view of recent investigation results on the foundations of aeroacoustics. In the process, the role and necessity of the so-called Stokes' hypothesis is reconsidered, which in its current practice omits the bulk viscosity. In contrast, the relevance of a new generalized kinematic viscosity is proposed, along with a new methodology that will enable estimation of second coefficient of viscosity completely obviating the necessity of making the Stokes' hypothesis. This opens up the vista of a new formulation for the Navier-Stokes equation applicable for both the subjects of fluid mechanics and aeroacoustics in a unified approach.

\end{abstract}



\section{Introduction}
The subject of acoustics involves study of acoustic signal or an acoustic disturbance consisting of pressure and density fluctuations traveling in an elastic medium. What makes the study of the propagation of acoustic signal very challenging is that the frequency and amplitude range of such signals variation is extremely wide. It is discussed in \cite{Sengupta_Bhumkar} that an young person can detect fluctuating pressure with amplitude as low as 20 $\mu Pa$, which is around ten orders of magnitude smaller compared to the background atmospheric pressure. It has also been discussed that the corresponding relative density fluctuation in a gaseous medium is given by, $\rho' / \rho_0 \approx M_0^2$, where $M_0$ and $\rho_0$ are the Mach number and density of the ambient fluid flow, respectively. It is for this reason, the study of acoustic signal propagation requires the study of compressible fluid flow as mandatory. Also, it is assumed that associated temperature change with acoustic signal propagation being negligible, the study invokes conservation of mass, momentum and an equation of state only for the five unknowns,  \cite{Schlichting}. 

It is necessary to broadly categorize computational aeroacoustics into the following methodologies. First, if one looks at free-field acoustics propagation as an inviscid phenomenon, then starting from Euler equation, one obtains the perturbation pressure field ($p'$) to be governed by the following as given in \cite{Feynman, Sengupta_Bhumkar},
$$\frac{\partial^2 p'}{\partial t^2} = c_0^2 \nabla^2 p'$$
Secondly, the perturbation pressure field by the so-called acoustic analogy field in the presence of mass source (G), non-conservative body force ($\vec{F})$, for a turbulent background flow as given in \cite{Lighthill, Ffowcs_Williams_Hawkins, Curle, Kinsler} is governed by, 
$$\frac{1}{c_0^2}\frac{\partial^2 p'}{\partial t^2} - \nabla^2 p' = \frac{\partial G}{\partial t} - \nabla\cdot \vec{F} + \frac{\partial^2 T_{ij}}{\partial x_i \partial x_j}$$
\noindent  where $T_{ij} = \rho u'_i u'_j + (p' - c_0^2\rho') \delta_{ij} - \tau_{ij}$ is the Lighthill's stress tensor. It is noted that the presence of mass source or the presence of body force, turbulent and viscous fluctuations converts the classical wave equation given in the left hand side to an inhomogeneous equations. In the literature, presence of mass source term on the right hand side is equivalent to having a monopole; the body force term is equivalent to having a dipole and the Lighthill's stress tensor causes a quadruple source to be present. The increased physical complexity due to these inhomogeneous terms can be difficult to interpret in explaining attenuation, dispersion, reflection, refraction, directionality and diffraction often noted even for simpler geometries for different flow conditions.  Finally, these difficulties can be circumvented by solving the set of compressible flow equations using direct numerical simulation (DNS). 

The central aspect of DNS is to solve the Navier-Stokes equation (NSE) resolving all the necessary length and time scales. In the present context of studying the interplay of acoustics and fluid flow, the problem of Rayleigh-Taylor instability (RTI) is an apt test to solve. While the topic of RTI is relevant from the atomic scale of inertial confinement fusion in \cite{Zhou_ARFM24} to astrophysical scale of supernovae in \cite{Cabot_Cook06}, here the attention is on a lab scale set up, with a miscible heavy fluid resting on top of a lighter layer of the same fluid, with experimental demonstration provided by \cite{Read} and a set of two-dimensional (2D) and three-dimensional (3D) DNS results reported in \cite{Sengupta_etal16POF} and \cite{Sengupta_etal22POF}, respectively. \cite{Sengupta_etal22POF} provides high  resolution DNS results that show the onset from the quiescent condition of heavier fluid resting on top of lighter fluid in a closed box to be associated with pressure pulse propagating from the interface, like planar one-dimensional (1D) wave-fronts. 

It is noted that there is the unresolved issue of the roles played by the second coefficient of viscosity ($\lambda$), which is a subject of many investigations. This is rooted in the Stokes' hypothesis, where \cite{Stokes} proposed vanishing of the bulk viscosity ($\mu_b = \lambda + \frac{2}{3} \mu $) for compressible flows for the sake of closure, with $\mu$ as the dynamic viscosity or the shear viscosity. 

Some discussions pertaining to Stokes' hypothesis is available in \cite{Sengupta_etal16POF}, and a quantitative estimate is provided in \cite{Sengupta_etal21CAF}, where the authors noted the effects of Stokes’ hypothesis on the onset of RTI and the growth of mixing layer at the incipient stage to have a strong dependence on whether Stokes' hypothesis is used or not. Computations reported by \cite{Sengupta_etal21CAF} without the Stokes' hypothesis, incorporated the non-zero values of bulk viscosity ($\mu_b$) obtained by a regression analysis of the acoustic attenuation and  dispersion measurements by \cite{Ash91, Zuckerwar} by solving the 3D compressible NSE. 
The small-scale billowing motion is only observed for non-zero bulk viscosity simulations. Following the onset, {\it the growth rates for bubbles and spikes in the mixing layer are found to be under-predicted by 12\% with the use of Stokes’ hypothesis}. Interested readers are referred to the  discussions on Stokes' hypothesis by \cite{Rosenhead, Gad_el_Hak, Rajagopal, Buresti}.  

There are certain difficulties and puzzling consequences if one accepts the current approaches for the use of Stokes' hypothesis. The foremost is that the bulk viscosity is eliminated by equating the mechanical and thermodynamic pressure, so as to relate $\lambda$ with $\mu$. However, any coefficient of viscosity implies loss of energy, and thus, cannot be negative. However with the Stokes' hypothesis, $\lambda = -2\mu/3$, the second coefficient of viscosity becomes anti-diffusive for the NSE, implying an inherent tendency of physical instability, which is improbable. As acoustic signal propagates as longitudinal waves, the successive compression and dilatation must result in loss of energy, and thus, $\lambda$ must be a positive diffusive coefficient. The second indirect consequence of Stokes' hypothesis is that the NSE involves the diffusion to be related solely to the kinematic viscosity ($\nu$). If one does not make use of the Stokes' hypothesis, then the governing NSE need not have only $\nu$ for the diffusion coefficient.

The above observation of dispensing with the Stokes' hypothesis is aided by our recent research in study of pulse propagation in a quiescent ambience, as was noted in the DNS reported in \cite{Sengupta_etal22POF}. Starting from the first principles, the governing equation for the perturbation pressure has been obtained in \cite{Sengupta_etal23POF} for multi-dimensional propagation that includes the effects of attenuation and dispersion along with its analytical solution. It is noted that there are classical textbooks which describe effects of viscous losses and heat conduction, as given in \cite{Morse, Trusler, Blackstock}. The present research aims at integrating these findings in deriving an appropriate governing equation for compressible flows, specifically a novel momentum conservation equation.

The paper is formatted in the following manner. In the next section, the perturbation pressure equation is described concisely, introducing a new generalized kinematic viscosity ($\nu_l = (\lambda + 2\mu)/\rho_0$) that incorporates both the dynamic viscosity coefficient and the second coefficient of viscosity. The properties of this governing equation are explained in terms of classifying the partial differential equation (following \cite{Ames}) in an appropriate parameter space. Section 3 describes the derivation of a new framework for the novel Navier-Stokes equation involving the presence of $\nu$ and $\nu_l$. In section 4, we propose a method for estimating $\nu_l$, either experimentally or numerically by following the planar acoustic pulse propagation. The paper closes with a summary and conclusion of the present investigation.

\section{Pressure perturbation equation and its properties}
\label{PPE}
As noted earlier, for the acoustic signal propagation, if the associated temperature change is negligible, then only the conservation of mass and momentum equation need to be considered, along with the equation of state, for the three unknown components of the velocity ($\vec{V}$), density and the pressure ($p$) as unknowns,

The conservation of mass is given by,

\begin{equation}
    \frac{\partial \rho}{\partial t} + \nabla \cdot (\rho \vec{V}) = 0 
    \label{Mass}
\end{equation}

The conservation of momentum equation for a compressible flow, without any body force is given by,

\begin{equation}
\rho (\frac{\partial \vec{V}}{\partial t} + (\vec{V}\cdot \nabla)\vec{V})= -\nabla p + \nabla \cdot (\lambda(\nabla \cdot \vec{V}) I) + \nabla \cdot [\mu(\nabla \vec{V} + \nabla \vec{V}^T)]
\label{Momentum}
\end{equation}

\vspace{3mm}

Here, $I$ is an identity matrix of rank three. 
For the perturbation analysis, let us express the unknowns as a sum of unperturbed state along with the perturbation quantities given by, 
$\vec{V} =\vec{\bar{V}} + \epsilon \vec{V}';\; \rho = \rho_0 + \epsilon \rho';\; p = \bar{p} + \epsilon p'$. 
Considering the propagation of acoustic signal in a quiescent ambience flow $(\vec{\bar{V}} =0)$ in a homogeneous medium ($\rho_0 = constant$), the perturbation equation from the conservation of mass equation yields,

\begin{equation}
\frac{\partial \rho'}{\partial t} + \rho_0  \nabla \cdot \vec{V}'= 0
\label{PMass}
\end{equation}

Similarly, the perturbation momentum equation can be derived from equation \eqref{Momentum} is obtained as,

\begin{equation}
\rho_0 \frac{\partial \vec{V}'}{\partial t} = -\nabla p' + \nabla \cdot (\lambda(\nabla \cdot \vec{V}') I) + \nabla \cdot [\mu(\nabla \vec{V}' + \nabla \vec{V}'^T)]
\label{PMomentum}
\end{equation}

Differentiating equation \eqref{PMass} with respect to time and taking the divergence of equation \eqref{PMomentum} enables one to eliminate perturbation velocity and perturbation density to write down the governing perturbation equation for pressure as,

\begin{equation}
\frac{\partial^2 p'}{\partial t^2} = c^2 \nabla^2 p' + \nu_l \frac{\partial}{\partial t} \nabla^2 p'
\label{PPressure}
\end{equation}

\noindent where we have introduced the generalized kinematic viscosity in \cite{Sengupta_etal23POF} as $\nu_l = \frac{\lambda +2\mu}{\rho_0}$, that incorporates both the dynamic viscosity and the second coefficient of viscosity. It is noteworthy that in deriving this equation, Stokes' hypothesis has not been invoked. The only assumption that is made is that the perturbation temperature is neglected during the propagation of the disturbance field and the perturbation density and perturbation pressure are related by the polytropic relation, $dp' = c^2 d\rho'$. It is to be noted that acoustic signal propagation is not an isentropic process. The actual polytropic process relates the perturbation pressure with the density perturbation as, $p' = K_1 (\rho')^n$, with $n$ as the polytropic index. Readers' attention is drawn to the elaborate discussion in \cite{PLA}, showing that one should replace $c^2$ by $nc^2/\gamma$; but the difference between $n$ and $\gamma$ is so marginal that one can neglect it as a second order effect. The presence of the viscous diffusion term in equation \eqref{PPressure}, makes the system dispersive, making the speed of sound length and time scale dependent. Thus, we do not require a constant speed of sound for all scales, as in classical wave equation and correspondingly one assumes the disturbance propagation as an adiabatic process. 

A major observation can be drawn from equation \eqref{PPressure}, which shows that in deriving the pressure perturbation equation, one does not require to make the Stokes' hypothesis, and the viscous dissipation is determined by $\nu_l$ and not $\nu$! One notices that in the textbooks, description of the NSE is given in terms of $\nu$, due to the custom of invoking the Stokes' hypothesis, whereby one equates the bulk viscosity to zero, to ensure the conjecture of equalizing mechanical pressure with the thermodynamic pressure. This helps eliminate the second coefficient of viscosity.  One of the strongest critiques of the use of the NSE for realistic acoustic wave propagation is due to the fact that sound attenuation is not due to shear viscous action only, but also due to the action of the bulk viscosity. This is due to the nature of acoustic signal that propagates as a longitudinal wave associated with compression and dilatation of the medium. We have already emphasized that the Stokes' hypothesis gives: $\lambda = -2\mu/3$, a completely unphysical result, because if shear viscosity is dissipative with $\mu > 0$, then the action of the second coefficient of viscosity will be physically unstable, as the corresponding $\lambda < 0$, indicates this as anti-diffusive!

Thus, it is imperative to develop a new common framework in developing the conservation of momentum equation for compressible flow without any need to resort to Stokes' hypothesis. This is provided in the next section, with the help of $\nu$ and $\nu_l$.

\section{Derivation of the Navier-Stokes equation without Stokes' hypothesis}

For the ease of understanding, we adopt the Cartesian frame of reference, starting with the Cauchy's equation given in the indicial notation (without summation implied for the repeated indices) for the conservation of momentum equation, as given in \cite{Schlichting, IFTT},



\begin{equation}
\rho \frac{Dv_k}{Dt} = X_k + \frac{\partial \tau_{ki}}{\partial x_i} + \frac{\partial \tau_{kj}}{\partial x_j} + \frac{\partial \sigma_{k}}{\partial x_k},\;\; {\rm for}\;\; k = 1,2,3
\label{z_cauchy}
\end{equation}

\noindent where the left hand side indicate the substantive derivative of the velocity components. Here, the velocity vector is written as, $\vec{V} = v_1 \hat{i} + v_2 \hat{j} + v_3 \hat{k}$, and the body force is written as, $\vec{F}_b = X_1 \hat{i} + X_2 \hat{j} + X_3 \hat{k}$, constituting the first terms on the right hand side of the above equation. The other stress gradient terms on the right hand side are due to surface forces. In writing the above surface force terms, one only considers isotropic, Newtonian fluid with linear constitutive relation between the state of stress with the rate of strain.

To secure isotropy and symmetry, the relationship between stress and rates of strain tensors can be directly evaluated as given in \cite{Lamb} or by using tensor algebra in \cite{Prager, Aris} to be given by, 
$$\sigma_x = -p + \lambda \nabla \cdot \vec{V} + 2\mu \frac{\partial v_1}{\partial x}$$
$$\sigma_y = -p + \lambda \nabla \cdot \vec{V} + 2\mu \frac{\partial v_2}{\partial y}$$
$$\sigma_z = -p + \lambda \nabla \cdot \vec{V} + 2\mu \frac{\partial v_3}{\partial z}$$
$$\tau_{xy} = \tau_{yx} =\mu \big(\frac{\partial v_2}{\partial x} + \frac{\partial v_1}{\partial y}\big)$$
$$\tau_{zy} = \tau_{yz} =\mu \big(\frac{\partial v_2}{\partial z} + \frac{\partial v_3}{\partial y}\big)$$
$$\tau_{xz} = \tau_{zx} =\mu \big(\frac{\partial v_3}{\partial x} + \frac{\partial v_1}{\partial z}\big)$$

\noindent where $\nu_l = \frac{\lambda}{\rho_0} + 2\nu$, and one can write down conservation of compressible momentum equation in indicial notation (with summation implied for repeated indices) as,

\begin{equation}
    \rho \big(\frac{\partial v_i}{\partial t} + v_j \frac{\partial v_i}{\partial x_j}\big) = X_i -\frac{\partial p}{\partial x_i} + \frac{\partial}{\partial x_j} \big\{ \rho \nu \big(\frac{\partial v_i}{\partial x_j} + \frac{\partial v_j}{\partial x_i}\big) + \rho(\nu_l -2\nu) \frac{\partial v_k}{\partial x_k} \delta_{ij}\big\} 
    \;\; (i,j,k = 1,2,3)
    \label{NNSE}
\end{equation}

\noindent where $\delta_{ij}$ is the Kronecker delta. Note that the conservation of mass remains unaltered, as given in equation \eqref{Mass}.

\section{Estimation of Generalized Kinematic Viscosity Coefficient ($\nu_l$)}
The significance of the perturbation pressure equation given in multi-dimensional form in equation \eqref{PPressure}, have been investigated in its one-dimensional form in \cite{Sengupta_etal23POF} with the governing equation given as,

\begin{equation}
    \frac{\partial^2p'}{\partial t^2} - c^2 \frac{\partial^2p'}{\partial x^2} - \nu_l \frac{\partial^3p'}{\partial t \partial x^2} = 0
    \label{PPE_1D}
\end{equation}

This is not a mere pedagogic exercise, as it has been clearly demonstrated in reporting the 3D DNS results in \cite{Sengupta_etal22POF} that the onset of RTI from quiescent condition, clearly demonstrated the receptivity route, by 1D excitation of pressure pulses spanning from infrasonic to ultrasonic band of frequencies, with different frequencies propagating the signal at different speed, as explained here due to dispersive, dissipative nature of pressure pulse propagation. Without going into the details, it suffices to note the receptivity route is a combination of modal and nonmodal routes, as has been noted in the literature, \cite{nonmodal_PRR, Sengupta21}. One can note this from the pressure spectra shown in figures 6 and 7 of \cite{Sengupta_etal22POF}.

If one represents the perturbation pressure in the spectral plane for 1D propagation as,
$$p'(x,t)= \iint \hat{p}(k,\omega)\;dk\; d\omega$$ \noindent then the dispersion relation is given by,

\begin{equation}
\omega^2 + i\nu_l k^2 \omega - c^2 k^2 = 0
\label{DR_1D}
\end{equation}
\noindent with the two components given by,
$$\omega_{1,2} = - \frac{i\nu_l k^2}{2} + \pm kc f$$
\noindent with $f = \sqrt{1-(\frac{\nu_l k}{2c})^2}$, and one notes that only ignoring the viscous action, one recovers the dispersion relation of the classical wave equation. It was noted \cite{Sengupta_etal23POF} that the wave (hyperbolic) nature of the perturbation pressure equation is lost if $f$ becomes purely imaginary, and one instead notices the governing equation to be purely parabolic partial differential equation for those high wavenumbers above $k_c (=2c/\nu_l)$. Such a wavenumber dependence on $\lambda$  is via $\nu_l$. This theoretical analysis in the spectral plane is by the well-known global spectral analysis (GSA), originally derived for numerical methods, and is reviewed in recent times by \cite{GSA_2023}. Its theoretical application on computational aeroacoustics are as in \cite{Sengupta_etal23POF, PLA}, where one can obtain the amplification factors, physical phase speeds and physical group velocity components with the value of $\lambda$ taken from the experimental results of \cite{Ash91}. 

\begin{figure*}
\centering
\includegraphics[width=.8\textwidth]{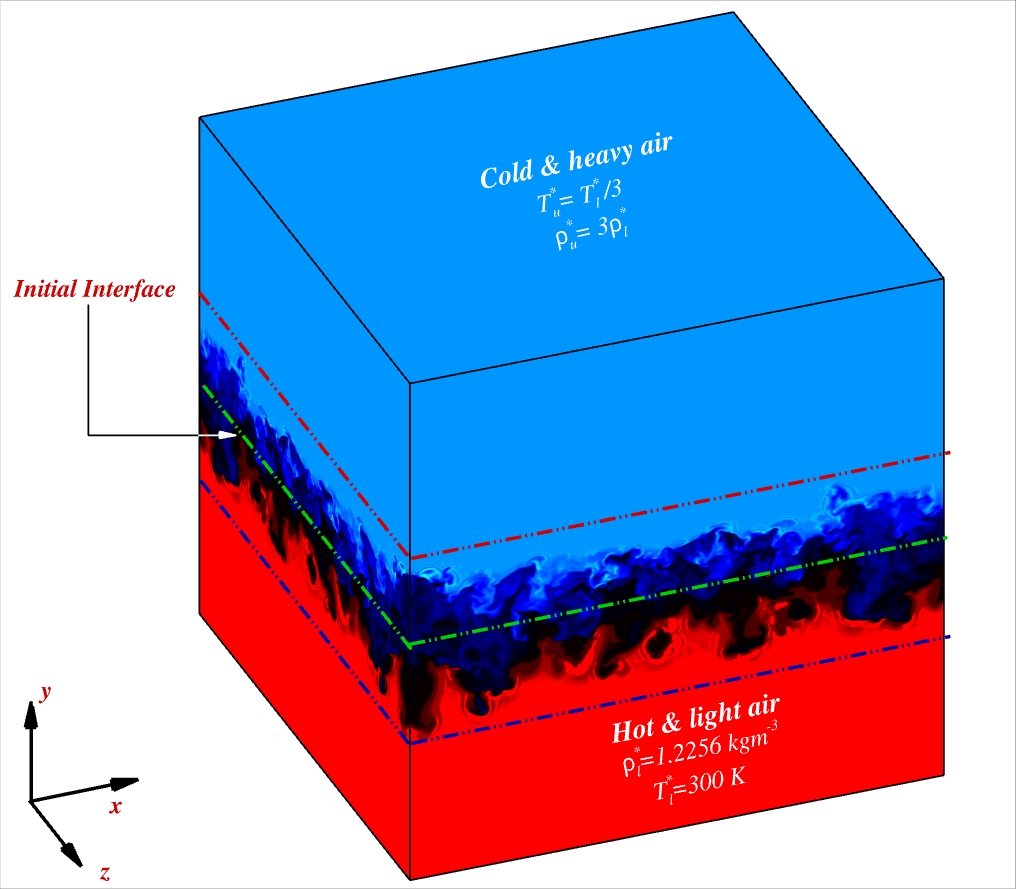}
\caption{Schematic of the RTI solved inside a cube of side $L = 0.15$m and original interface at $y^* = L/2$.}
\label{fig1}
\end{figure*}

The use of 1D study is also due to the fact that the dispersion relation for any 3D field is given by a sixth order polynomial, which is not solvable analytically, while the 1D field is amenable for analysis. Even though, one can analyze the perturbation pressure field by solving the quartic dispersion relation equation in 2D,  our attention is focused to 1D pressure pulse propagation, as in our previous researches on the effects of bulk viscosity for RTI in 2D (\cite{Sengupta_etal16POF}) and 3D flows (\cite{Sengupta_etal22POF}), have revealed the onset of RTI is by acoustic excitation from the quiescent state by 1D pressure pulses. The case of 3D DNS of RTI in \cite{Sengupta_etal22POF} is revisited here in explaining the role of $\nu_l$ in unifying fluid mechanics and aeroacoustics via 1D pressure pulse propagation.

Having noted that the pressure perturbation equation is developed for a quiescent ambience, the choice of the RTI problem is natural, whose schematic is shown in figure \ref{fig1} for the investigated problem with details in \cite{Sengupta_etal22POF, PLA}, providing the governing equations, supplemented by the equation of state for perfect gas in non-dimensionalized form. Briefly, the description of the problem is given here in the following.

The problem consists of air in a cube of physical dimension of $L = 0.15m$ into two compartments kept at dissimilar temperature, and separated by an impermeable, insulating diaphragm. The statically unstable initial configuration consists of the cold air (at $100K$) kept on top, and the relatively hotter air in the bottom compartment (at $300K$). At $t= 0$, the diaphragm is removed impulsively, and that gives rise to acoustic pulses whose spectral content extends beyond ultrasound limit. The velocity and time scales are chosen as 12.131m/s and 0.1237s, respectively. The viscosity scale is taken as the dynamic viscosity of air at $T_s = 300K$. For the DNS database created in \cite{Sengupta_etal22POF, Sundaram_etal22_JCP} is for the following non-dimensional parameters: Reynolds number of 12,080.6; Atwood number of 0.5; Prandtl number of 0.712, Froude number of 1 and Mach number of $3.4939 \times 10^{-3}$. The cubic domain is solved by the compact schemes with fourth order, four stage Runge-Kutta time integration method using ($1280 \times 2560 \times 1280$) grid points in $x$-, $y$- and $z$-directions, respectively. The simulation resolves up to 6.5019MHz, with time steps chosen as $7.69 \times 10^{-8}s$. The bulk viscosity is taken by a regression formula of the experimental data provided in \cite{Ash91}, which fixes $\nu_l$, along with the Sutherland's formula for the variation of $\mu$ with temperature. 

\begin{figure*}
\centering
\includegraphics[width=\textwidth]{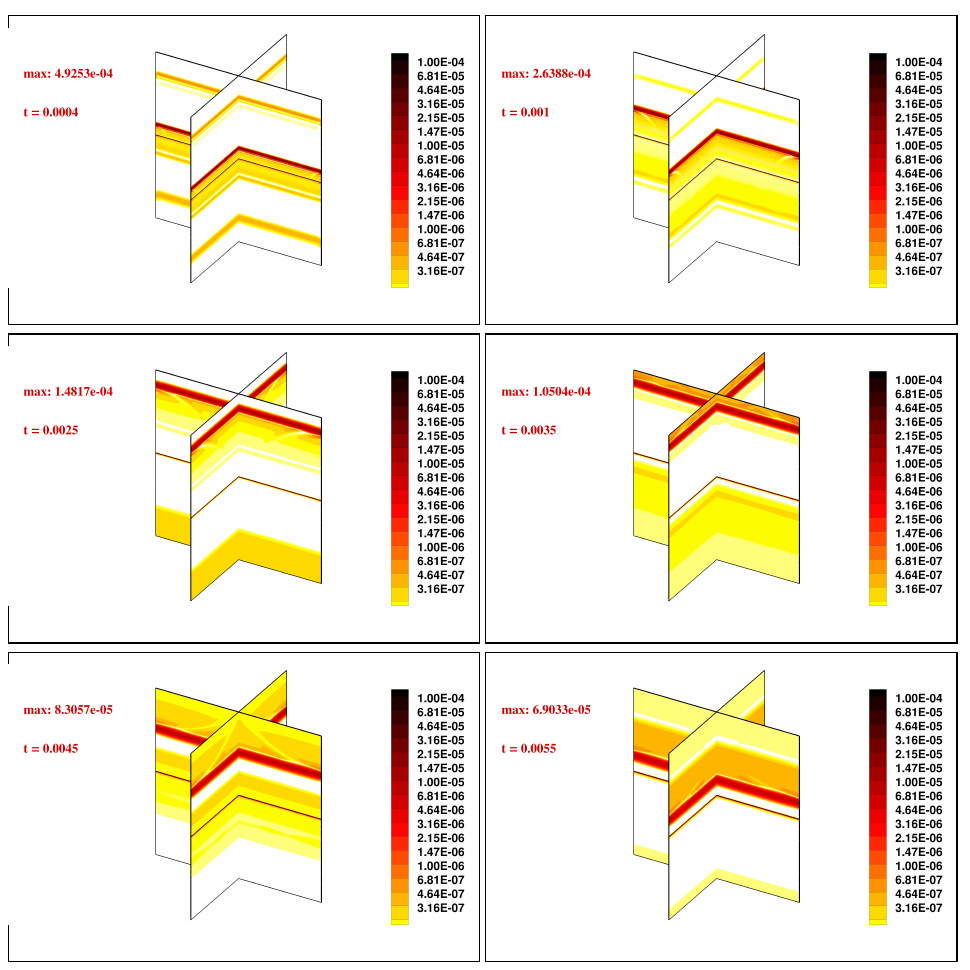}
\caption{Schematic of the RTI solved inside a cube of side $L = 0.15$m and original interface at $y^* = L/2$.}
\label{fig2}
\end{figure*}

The disturbance pressure is obtained by subtracting the initial static pressure from the instantaneous one, and the typical snapshots for this disturbance pressure are plotted along $y$-direction passing through ($z* = x* = L/2$) in figure \ref{fig2}. The snapshots are for very early time instant events, when the flow field is virtually at quiescent state. As $p'$ evolves showing iso-contours as parallel lines, perpendicular to the normal of the initial interface, this is clearly a 1D propagation of the disturbance pressure field, with the pulses propagating along $y$-direction, substantiating the 1D analysis by GSA of the results.  

\begin{figure*}
\centering
\includegraphics[width=\textwidth]{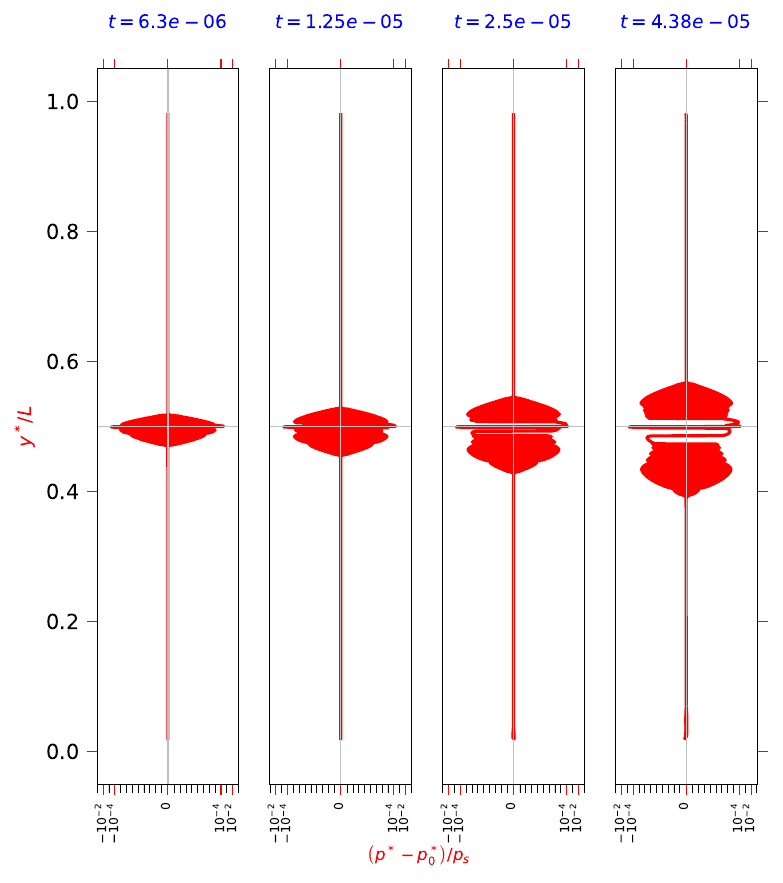}
\caption{Disturbance pressure plotted as a function of $y$ at early instants indicated in the labels. Here, $p^*$ is the instantaneous pressure, $p^*_0$ is the initial pressure, and $p_s$ is pressure scale corresponding to hot air.}
\label{fig3}
\end{figure*}

The pressure pulses caused by impulsively removing the interface  (across which the temperature variation follows a Heaviside function), displays 1D variation of $p'$ with $y$ at the onset, even though the RTI is essentially a 3D problem. The variation of $p'$ versus $y$ is shown in figure \ref{fig3}, along the vertical centerline of the cubic domain, originating from the interface, propagating in the opposite direction from the interface. In \cite{Sengupta_etal22POF}, the detailed results are depicted, along with the Fourier transform of $p'$ that indicates the dispersive nature of the field. It is according to the GSA of the 1D $p'$ propagation of modal and nonmodal variations, as shown in \cite{Sundaram_etal22_JCP, Sengupta_etal22POF}. 

Despite the clear 1D dynamics of $p'$ propagation during the onset of RTI, such an experimental set-up is not ideal for estimating $\nu_l$ due to the rich pressure spectrum noted, spanning across infrasonic to ultrasonic frequencies. Instead, one can follow the pulse  propagation described in \cite{PLA}, where the theoretical set-up consists of a time-impulsive Gaussian wave-packet created at $t=0$ given by the initial conditions,
$$p'(x,0) = e^{(-\alpha (x - x_0))} \sin (k_0 (x-x_0))$$
\noindent where $x_0 =0$ and $\alpha = 500$, and $k_0$ is the central wavenumber of the Gaussian pulse. The requisite second initial condition used is given by,
$$\frac{\partial p'}{\partial t} (x,0) = 0$$

In \cite{PLA}, the detailed numerical simulation of the 1D pressure pulse equation is reported using the compact scheme in \cite{Lele} for the second spatial derivative, along with the four stage, fourth order Runge-Kutta time integration scheme in solving equation \eqref{PPE_1D}. Representative property charts have been shown for four different values of $\nu_l$. However, the pulse propagation solutions of equation \eqref{PPE_1D} have been shown only for the case of $\nu_l = 0.1443m^2/s$, which corresponds to the experimental results for the estimated $\lambda$ in terms of $\mu$ for air in \cite{Ash91}. To the best of author's knowledge, there are no other experimental results for the second coefficient of viscosity for air. There is an urgent need to estimate $\nu_l$ for common fluids like air and water at varying ambient conditions. 

Now that the correct pressure perturbation equation has been theoretically proposed in \cite{Sengupta_etal23POF, PLA}, and with the help of which the recasting of the Navier-Stokes equation has been performed here in terms of $\nu$ and $\nu_l$, the task that remains is the evaluation of $\lambda$ experimentally or theoretically. The experiment to achieve that can be the equivalent of the set-up of the physical scenario reported in \cite{PLA} for 1D pressure pulse propagation. The experimentally obtained $\lambda$ and $\nu_l$ can be cross checked and validated via the highly accurate simulation of the equation \eqref{PPE_1D}. There is also, the possibility to refine the estimate of $\nu_l$ via the usage of estimation theory (as in extended Kalman filter) and/ or using machine learning tools.

\section{Summary and Conclusions}
     The role of Stokes hypothesis is critically discussed, and some of its shortcomings due to this are highlighted. One of the strongest critiques and an unacceptable aspect of the Stokes' hypothesis is the role of the second coefficient of viscosity turning to anti-diffusive due to this hypothesis ($\lambda = -2\mu/3$) is emphasized.
     
    In another exercise on acoustics pulse propagation in quiescent ambience, the author's group has highlighted another aspect of the conservation of mass and momentum equation (as given by the Cauchy's equation), with the correct equation derived without the Stokes' hypothesis for the perturbation pressure. Most importantly, a new generalized kinematic coefficient of viscosity, $\nu_l\; [ = (\lambda + 2\mu)/\bar{\rho}$] is noted as the correct combination of the coefficients of viscosity that must be incorporated in the revised Navier-Stokes equation, as given by equations (2.1) and (3.2) for the conservation of mass and momentum. This observation does not invoke any additional conditions on conservation of energy equation, which may have to be also solved for the combined problem of fluid mechanics and aeroacoustics. This is the approach adopted in \cite{Sengupta_etal22POF} for the RTI problem, providing acoustic and fluid dynamic variables as the imprint of the solutions.
    
    The major issue remains is to estimate $\nu_l$. Experimental measurements of $\lambda$ are practically non-existent, except that is due to \cite{Ash91, Zuckerwar}. This value of $\lambda$ was used by \cite{Sengupta_etal22POF} for the 3D DNS of Rayleigh-Taylor instability (RTI), and two features were noted from this extremely high accuracy HPC results obtained by compact scheme using more than four billion grid points in a cube: (i) At the onset of the RTI when the flow inside the cubic domain was virtually quiescent, one noted one-dimensional pressure pulses emanated perpendicular to the initial unperturbed interface; (ii) The spectrum of the pressure field demonstrated both modal and nonmodal disturbance propagation with frequencies spanning from infrasonic to ultrasonic ranges.  
    
    The presence of 1D pressure pulse propagation was theoretically analyzed using GSA in \cite{Sengupta_etal23POF, PLA}. However, the complex spectrum of pressure perturbation during RTI is not amenable in estimating $\lambda$.
    
     Instead, 1D pressure pulse propagation caused by impulsively started Gaussian wave-packet with a single wavenumber was numerically simulated in \cite{PLA}. This was simulated with the value of $\lambda$ measured by \cite{Ash91}. This is proposed here finally to estimate $\lambda$ via experimentation and ML methods. 



\backsection[Funding]{This research received no specific grant from any funding agency, commercial or not-for-profit sectors.}

\backsection[Declaration of interests]{The authors report no conflict of interest.}

\bibliographystyle{jfm}
\bibliography{ROE}

\end{document}